\documentclass[preprint,showpacs,preprintnumbers,amsmath,amssymb]{revtex4}

\usepackage{graphicx}
\usepackage{dcolumn}
\usepackage{bm}

\begin{document}

\preprint{APS/123-QED}

\title{Theory of Fluctuations in a Network of Parallel Superconducting Wires}

\author{Kohjiro Kobayashi and David Stroud}
\affiliation{%
Department of Physics, Ohio State University, Columbus, OH 43210\\
}%

\date{\today}

\begin{abstract}
We show how the partition function of a network of parallel superconducting wires weakly coupled together by the proximity effect, subjected a vector potential along the wires can be mapped onto N-distinguishable two dimensional quantum-mechanics problem with a perpendicular imaginary magnetic field. Then, we show, using a mean field approximation, that, for a given coupling, there is a critical temperature for onset of inter-wire phase coherence. The transition temperature $T_c$ is plotted on both cases for non-magnetic and a magnetic field perpendicular to the wires.
\end{abstract}

\maketitle

\section{Introduction}
There has been considerable recent interest in thin wires that undergo transitions into an ordered state, such as superconducting or ferromagnetic. For example, a recent experiment \cite{Tang2001} has suggested that single-walled carbon nanotubes (which have diameters of only about 4 \AA) are superconducting up to temperatures as high as 20 K. Because these tubes are so thin, they behave very much like one-dimensional superconductors. It was therefore proposed \cite{Tang2001} that they could be described by a complex order parameter $\psi(z)$ which varies only in one dimension, say the z direction, i.e. along the tube. $\psi(z)$ might represent the complex energy gap, or, in a different normalization, it could represent the condensate wave function in a BCS superconductor.

Moreover, there have been many experiments for investigating superconductivity on nanowires. Ropes of carbon nanotubes between superconducting electrodes can show superconductivity due to the proximity effect of the electrodes \cite{Kasumov1999, Morpurgo1999, Gonzalez2001}. Furthermore, superconductivity on carbon nanowires connected to normal contacts, has been observed \cite{Kociak2001, Kasumov2003}. On the other hand, superconductivity of nanowires of Zn or Sn has been investigated \cite{Tian2005a, Tian2005b}.

Fluctuations are, of course, especially important in one-dimensional systems. It was shown many years ago by Scalapino {\it et al} \cite{Scalapino1972} that {\em classical} fluctuations in one dimension could be treated {\em exactly}, within the context of a Ginzburg-Landau (GL) free energy functional. Their treatment involved mapping the GL functional onto a single-particle quantum mechanics problem, using an exact connection between the classical partition function and a path integral treatment of the quantum mechanics problem. These authors showed that classical fluctuations could give rise to a non-zero order parameter even above the GL transition temperature. This mapping was extended to treat Josephson-coupled thin wires \cite{Stoeckly1975, Scalapino1975}.

However, in the mapping, the effect of a magnetic field was ignored. In the case of a non-zero perpendicular magnetic field, we show that the partition function for the wires maps onto a certain zero-temperature quantum mechanics problem in two dimensions with an effective imaginary perpendicular magnetic field, which brings to a non-Hermitian quantum mechanics problem.

The non-Hermitian problem in physics has not been new recently. Nonequilibrium processes can be described by non-Hermitian Liouville operators \cite{Kadanoff1968, Fogedby1995, Kim1995}. The non-Hermitian quantum mechanics are well studied in order to study the pinning of magnetic flux lines in high temperature superconductors \cite{Nelson1993, Hatano1996, Hatano1997}.

The remainder of this paper is organized as follows. In Section \ref{sec:one_formalism}, we describe our formalism and mapping. In Section \ref{sec:one_result}, we give our numerical results including phase diagrams. This is followed by a concluding discussion and an outline of possible future research.

\section{Formalism \label{sec:one_formalism}}
\subsection{Mapping to a quantum mechanics problem for interacting
superconducting wires when $ \vec{B} $ is perpendicular to the wires} Let us consider a network of N parallel superconducting wires in a non-zero vector potential. We assume, for convenience, that these wires all have the same GL parameters, though the formalism can easily be generalized to the case when the parameters are different. Then the partition function can be written as a functional integral over the $N$ complex order parameters $\psi_1(z_1),...\psi_N(z_N)$:
\begin{equation}
Z = \int{\cal D}\psi_1(z_1)...{\cal D}\psi_N(z_N)\exp\{-\beta
F[\psi_1(z_1),...\psi_N(z_N)]\}.
\end{equation}
We assume that the free energy functional is the sum of two parts: a single-wire term $F_s$ and a term describing inter-wire interactions, which we denote $F_{int}$. The single-wire term will just be the sum of general GL equation for each wire:
\begin{equation}
F_s = \sum_{i=1}^NF_{GL}[\psi_i(z_i)].
\end{equation}
Here,
\begin{equation}
F_{GL}[\psi_i(z_i)] =
\int_0^{z_{max}}[\frac{1}{2m^*}|\left(\frac{\hbar}{i}\nabla-\frac{e^*\vec{A}
}{c} \right)\psi (z)|^2+\alpha |\psi(z)|^2+\gamma
|\psi(z)|^4+\frac{HB}{8\pi}\Sigma]dz,
\end{equation}
where $\alpha$, $\gamma$, and $m^*$ are material-dependent (and possibly temperature-dependent) coefficients. Commonly, it is assumed that $\gamma$ is positive and that $\alpha = \alpha^\prime(T - T_c)$, where $T$ is the temperature, $T_c$ is the critical temperature, and $\alpha^\prime$ is greater than zero. Also, $\Sigma$ is the cross-sectional area of the sample, but for one-dimensional wire we may ignore this term. For the interaction term, we assume a form similar to that used by Lawrence and Doniach for interacting superconducting layers \cite{Lawrence}, namely
\begin{equation}
F_{int} = \sum_{\langle ij \rangle}\int_0^{z_{max}}K_{ij}|\psi_i(z)
- \psi_j(z)|^2.
\end{equation}
where $z_{max}$ is the length of the wires. Basically, we are assuming that there is a Josephson coupling of strength $K_{ij}$ between different wires, but at the same point along the length, $z$. We choose a gauge such that the vector potential is parallel to the superconducting wires, has only $z$ component and independent of $z$. When a wire is a loop, a vector potential is related to the total flux $\Phi$ through the loop, $A_z = \Phi /z_{max}$.
In this case, using
$\psi_i(z)=\psi_{iR}(z)+i\psi_{iI}(z)$, $F_s$ and $F_{int}$ take the forms
\begin{equation}
F_s \!=\! \sum_i\!\!\int_0^{z_{max}}\!\!\!\!\![\frac{\hbar^2}{2m^*}|\psi_i'|^2
-\frac{e^*\hbar A_z}{m^*c}(\psi_{iR}\psi_{iI}'-\psi_{iR}'\psi_{iI})
+\left\{\!\alpha+\frac{1}{2m^*}\left(\frac{e^*}{c}\right)^2A_z^2\!\right\}\!\!|\psi_i|
^2 + \gamma |\psi_i|^4]dz, \label{eq:sheng1}
\end{equation}
and
\begin{equation}
F_{int} = \sum_{\langle ij \rangle}\int_0^{z_{max}}K_{ij}
(|\psi_i(z)|^2+|\psi_j(z)|^2-2(\psi_{iR}\psi_{jR}+\psi_{iI}\psi_{jI})), \label{eq:sheng2}
\end{equation}
where $\psi '(z)=d\psi(z)/dz$. Finally, the partition function takes the form
\begin{equation}
Z = \int\sum_{i}{\cal D}\psi_{iR}{\cal D}\psi_{iI}\exp(-\beta
F[\psi_{iR},\psi_{iI} ]), \label{eq:sheng3}
\end{equation}
where we use $\psi_{i}=\psi_{iR}+i\psi_{iI}$.

We now show that eqs.\ (\ref{eq:sheng1}), (\ref{eq:sheng2}) and (\ref{eq:sheng3}) for $Z$ are actually equivalent to a {\em quantum mechanical} problem of a N distinguishable particles in N distinct quantum wells in two dimensions in the presence of a perpendicular magnetic field. In order to simplify our argument, we consider the case of single particle with mass, $m$ and a charge $e^*$ subjected to a 2D potential, $V(x,y)$. The density matrix of a two-dimensional system, using $\psi^I$ and $\psi^F$ are boundary condition at initial and final time, can be written \cite{Feynman1965}
\begin{equation}
\langle \psi^F|e^{-S/\hbar}|\psi^I\rangle
 = \int_{\psi^I}^{\psi^F}{\cal D}x(\tau){\cal D}y(\tau)\exp{\{-\frac{1}{\hbar}S[x(\tau),y(\tau)]\}},
\end{equation}
where
\begin{equation}
S =
\int_{0}^{\beta_{eff}\hbar}[\frac{m}{2}(x'^2+y'^2)+V(x,y)-i\frac{e^*}{c}\vec
{A}\cdot\vec{v}]d\tau .
\end{equation}
For the given $\vec{B}=B_{eff}\hat{z}$ with the gage
\begin{equation}
\vec{A}_{eff} = \frac{B_{eff}}{2}(x\hat{y}-y\hat{x}),
\end{equation}
this $S$ becomes
\begin{equation}
S =
\int_{0}^{\beta_{eff}\hbar}d\tau[\frac{m}{2}(x'^2+y'^2)+V(x,y)-i\frac{e^*B_{eff}}
{2c}(xy'-yx')] .
\end{equation}
This is a similar equation to the partition function of the superconducting wires.

In order to simplify this mapping, we use the suitable dimensionless form. $\tau = \frac{\beta\hbar}{\xi_0}z, \tilde{\psi}_{ix} = \xi_0^{3/2} \psi_{iR}$, and $\tilde{\psi}_{iy} = \xi_0^{3/2} \psi_{iI}$.
Then we can make the identifications of Table~\ref{tab:table1}.
\begin{table}
\begin{center}
\begin{tabular}{|c||c|}
\hline
 Q.M. & S.C. \\
\hline\hline
 $\tau$ & $z$\\
\hline
 $\vec{\rho_i}=\{x_i(u),y_i(u)\}$ & $\vec{\psi_i}=\{\tilde{\psi}_{ix}(z),\tilde{\psi}_{iy}(z)\}$ \\
\hline
 $E$ & $F\frac{\xi_0}{z_{max}}$ \\
\hline
 $\beta_{eff}$ & $\beta z_{max}/\xi_0$ \\
\hline
 $V_i(x_i,y_i)$ & $\left(\frac{\alpha}{\xi_0^2}+\frac{1}{2m^*\xi_0^2}(\frac{e^*A_z}{c})^2\right)|\tilde{\psi}_i|^
2+\frac{\gamma}{\xi_0^5}|\tilde{\psi}_i|^4$ \\
\hline
 $m$ & $\frac{\hbar^4\beta^2}{m^*\xi_0^4}$ \\
\hline
 $B_{eff}$ & $-i\frac{2\hbar^2A_z\beta }{m^*\xi_0^3}$ \\ 
\hline
 $J_{ij}$ & $\frac{K_{ij}}{\xi_0^2}$ \\
\hline
\end{tabular}
\caption{\label{tab:table1}Correspondence on the mapping between Q.M. and S.C. (in each case, the left-hand variable corresponds to the parameters on the quantum mechanics problem and the right hand variable corresponds to the parameters on the superconductor wires)}
\end{center}
\end{table}

We find that the magnetic field has two effects: (i) it determines an effective perpendicular magnetic field in which the equivalent quantum-mechanical particle moves; and (ii) it changes the quadratic part of the effective potential. The Hamiltonian for the analogous quantum problem is
\begin{equation}
H =
\sum_{i=1}^N[\frac{1}{2m}\left(p_{ix}+\frac{e^*B_{eff}}{2c}y\right)^2+\frac{1}{2m}\left(p_{iy}-\frac{e^*B_{eff}}{2c}x\right)^2
+V_i(\vec{\rho}_i)]+\sum_{<ij>}2J_{ij}|\vec{\rho}_i-\vec{\rho}_j|^2
,\label{eq:effh}
\end{equation}
where $p_{ix}$ and $p_{iy}$ are momentum operators of x and y components of $i$th particle, respectively.

\subsection{Probability distribution of the order parameter}
We consider the probability distribution of the order parameter, which corresponds to the probability distribution of particles in quantum mechanics. In order to simplify our discussion, we consider single wire case. The probability distribution function of the order parameter can be defined as
\begin{equation}
P(\vec{\rho}(\tau))=\frac{1}{Z}\langle\psi^F|e^{-\frac{H}{\hbar}(L_{\tau}-\tau)}|\vec{\rho}(\tau)\rangle\langle\vec{\rho}(\tau)|e^{-\frac{H}{\hbar}\tau}|\psi^I\rangle ,
\end{equation}
where $Z=\langle\psi^F|e^{-\frac{H}{\hbar}L_{\tau}}|\psi^I\rangle$ and $|\psi^I\rangle$ represents the boundary condition at $\tau=0$ and $\langle\psi^F|$ represents the boundary condition at $\tau=L_{\tau}$. Using the eigenstates of the Hamiltonian, $H|n\rangle=E_n|n\rangle$, the probability can be written as
\begin{equation}
P(\vec{\rho}(\tau))=\frac{1}{Z}\sum_{m,n}\langle\psi^F|m\rangle\langle m|\vec{\rho}(\tau)\rangle\langle\vec{\rho}(\tau)|n\rangle\langle n|\psi^I\rangle e^{-\frac{E_m}{\hbar}(L_{\tau}-\tau)}e^{-\frac{E_n}{\hbar}\tau}
\end{equation}
with
\begin{equation}
Z=\sum_n\langle\psi^F|n\rangle\langle n|\psi^I\rangle e^{-\frac{E_n}{\hbar}\tau} .
\end{equation}
Explicitly, the expectation value of operator, $\rho$ at the distance $\tau$ from the bottom of the wires is given by
\begin{equation}
\langle\hat{\rho}\rangle_{\tau}=\frac{1}{Z}\langle\psi^F|e^{-\frac{H}{\hbar}(L_{\tau}-\tau)}\int d\vec{\rho}(\tau)|\vec{\rho}(\tau)\rangle\rho\langle\vec{\rho}(\tau)|e^{-\frac{H}{\hbar}\tau}|\psi^I\rangle ,
\end{equation}
where $\hat{\rho}|\vec{\rho}\rangle=\rho|\vec{\rho}\rangle$.

The case of periodic boundary condition, our problem can be simplified. If $\psi^F$ corresponds to $\psi^I$ and summed over
all possible this configuration, the density matrix can be written as. 
\begin{eqnarray*}
P(\vec{\rho}(\tau))&=&\frac{1}{Z}\sum_{m,n}\sum_I\langle\psi^I|m\rangle\langle m|\vec{\rho}(\tau)\rangle\langle\vec{\rho}(\tau)|n\rangle\langle n|\psi^I\rangle e^{-\frac{E_m}{\hbar}(L_{\tau}-\tau)}e^{-\frac{E_n}{\hbar}\tau} \\
&=&\frac{1}{Z}\sum_{n}\langle n|\vec{\rho}(\tau)\rangle\langle\vec{\rho}(\tau)|n\rangle e^{-\frac{E_n}{\hbar}L_{\tau}} ,
\end{eqnarray*}
where $Z=\sum_n\sum_I\langle\psi^I|n\rangle\langle n|\psi^I\rangle e^{-\frac{E_n}{\hbar}L_{\tau}}=\sum_n e^{-\frac{E_n}{\hbar}L_{\tau}}$. So, if the wire is actually in the form of a loop, which means the boundary conditions $\psi(0)=\psi(z_{max})$, our problem corresponds to this statistical mechanics. Of course, in the limit of a very long wire, the periodic boundary condition imposed by the loop should become unimportant.

In the case of the periodic boundary condition for single wire, we can see qualitative behavior of order parameter. The average gap in the GL problem (denoted $\tilde{\Delta}(t)$) corresponds to the mean distance
$\langle \rho \rangle$ in the quantum-mechanical problem, i.e.
\begin{eqnarray}
\langle\rho\rangle \leftrightarrow \tilde{\Delta}(t) ,
\end{eqnarray}
where $\Delta (t) = \tilde{\Delta}(t)/\xi_0^{3/2}$. At much lower temperature than the critical temperature $T_c^0$, the mean distance from the origin of the particle approaches the value predicted for the quantum problem in the limit of infinite mass, i.e. the value of $\rho$ for which the quartic potential is a minimum although the magnitude of these gaps at $T=0$ are different. The function $\sqrt{1-t}$, is the classical solution, i.e., in the case when thermal fluctuations in the GL case are negligible. These fluctuations do indeed become very small when $T\rightarrow 0$, because in this regime, the effective potential rises steeply above its minimum, and the $\langle\rho\rangle$ becomes very close to the value that minimizes the GL free energy. When $\langle\rho\rangle$ has this value, the corresponding value for $\tilde{\Delta}(t)$ is
\begin{equation}
\tilde{\Delta}(t) = \sqrt{\tilde{\psi_R}^2+\tilde{\psi_I}^2} =
\sqrt{\frac{\alpha_0 T^0_c\xi_0^3}{2\gamma}}\sqrt{1-t}=\tilde{\Delta}(0)g(t), \label{eq:rho_one}
\end{equation}
where $\tilde{\Delta}(0)$ is the gap at $T=0$. These considerations may suggest that we can approximate $g(t)=\tilde{\Delta}(t)/\tilde{\Delta}(0)=\sqrt{1-t}$.

\subsection{Phase only model and mean-field approximation}
This system will undergo a phase transition into a phase-ordered state below a critical temperature $T_c$ which is distinct from (and lower than) the single wire mean-field transition temperature $T_{c}^0$. To do this, we consider a simplified, ``phase-only'' version of this Schr\"{o}dinger equation (\ref{eq:effh}). We assume that the {\em magnitudes} $\rho_i$ of the variables ${\bf x}_i$ are fixed at the values which minimize the single-wire GL free energy, i.e. $\rho_i \equiv \rho_0$ (\ref{eq:rho_one}). All terms in the Hamiltonian involving $\partial/\partial\rho_i$ can be ignored in this phase-only model. The effective Hamiltonian (\ref{eq:effh}) then becomes
\begin{equation}
H =
-\sum_i\frac{\hbar^2}{2m\rho_0^2}\frac{\partial^2}{\partial\phi_i^2}
-\sum_i\frac{e^*B_{eff}}{2mc}\frac{\hbar}{i}\frac{\partial}{\partial\phi_i}
+2\sum_{\langle ij\rangle}J_{ij}\rho_0^2(1-\cos{(\phi_i-\phi_j)}), \label{eq:phaseH}
\end{equation}
where the sum runs over distinct nearest neighbor pairs. This is the well-known {\em quantum XY model}, which exhibits a quantum phase transition at a critical value.

The mean field approximation can be applied to this Hamiltonian, assuming that $J_{ij}=J$ for only nearest neighbors, by replacing the second term according to the prescription
\begin{equation}
\cos{(\phi_i-\phi_j)}=2\cos{\phi_i}\langle\cos{\phi}\rangle-\langle\cos{\phi}\rangle^2 ,
\end{equation}
where we are supposing $\langle\sin{\phi}\rangle=0$ because of the symmetry.
Thus,
\begin{eqnarray*}
2\sum_{\langle ij\rangle}J\rho_0^2 (1-\cos{(\phi_i-\phi_j)})
&=&2\sum_{\langle ij\rangle}J\rho_0^2\left(1-2\cos{\phi_i}\langle\cos{\phi}\rangle+\langle\cos{\phi}\rangle^2\right) \\
&=&-4z_nJ\rho_0^2\langle\cos{\phi}\rangle\sum_i\cos{\phi_i}+2\sum_{\langle ij\rangle}J\rho_0^2(1+\langle\cos{\phi}\rangle^2) \\
&=&-4z_nJ\rho_0^2\langle\cos{\phi}\rangle\sum_i\cos{\phi_i}+2z_nNJ\rho_0^2(1+\langle\cos{\phi}\rangle^2) ,
\end{eqnarray*}
where $z_n$ is the number of nearest neighbors in the lattice.
Thus, the effective Hamiltonian corresponding to eq.\ (\ref{eq:phaseH}) becomes a following Schr\"{o}dinger equation:
\begin{equation}
\left\{\!\!\frac{-\hbar^2}{2m\rho_0^2}\frac{\partial^2}{\partial\phi_i^2}
\!\!-\!\frac{e^*B_{eff}}{2mc}\frac{\hbar}{i}\frac{\partial}{\partial\phi_i}
\!\!-\!4z_n\rho_0^2J\langle\cos{\phi}\rangle\!\!\cos{\phi_i}\!\!+\!2z_nJ\rho_0^2(1\!\!+\!\langle\cos{\phi}\rangle^2)
\!\!\right\}
\!\!\psi_n(\phi_i)\!
= \! E_n\psi_n(\phi_i). \label{eq:phaseonlyh}
\end{equation}

We consider the self consistent equation for $\cos{\phi}$ on the periodic boundary condition.
The mean field theory is defined by the self-consistency requirement
on $\langle\cos{\phi}\rangle$:
\begin{equation}
\langle\cos{\phi}\rangle =\frac{\sum_n
e^{-\beta_{eff}E_n}\langle\psi_n(\phi_i)|\cos{\phi_i}|\psi_n(\phi_i)\rangle}
{\sum_n e^{-\beta_{eff}E_n}}. \label{eq:condition}
\end{equation}
For example, when the wires are sufficiently long where only the ground state contribution may be important, the self-consistent condition becomes
\begin{equation}
\langle\cos{\phi}\rangle=\langle\psi_0(\phi_i)|\cos{\phi_i}|\psi_0(\phi_i)\rangle . \label{eq:gcondition}
\end{equation}
These equations may be solved for $\langle\cos{\phi}\rangle$ and $T_c$, where the critical temperature can be determined by $\langle\cos{\phi}\rangle\rightarrow 0$.

\section{Results and Discussion \label{sec:one_result}}
We have considered long-range phase coherence among wires in the bundle in order to see whether the phases on the wires are coherent and the bundle as a whole is superconducting or not. The self-consistent equation gives rise to a phase diagram exhibiting superconductivity, which can be defined as the greatest temperature and field such that $\cos{\theta}$ takes on a non-zero value \cite{Simanek1979}. Here, we assume that the Josephson coupling is independent of a temperature. We consider the temperature dependence, $\sqrt{1-t}$ for $\rho$. In order to simplify our calculations, we consider the case of the periodic boundary condition.

\subsection{No magnetic field}
We consider the following self-consistent equation, substituting $B_{eff}=0$ for the differential eq.\ (\ref{eq:phaseonlyh}),
\begin{equation}
\left(-\frac{\hbar^2}{2m\rho_0^2}\frac{\partial^2}{\partial\phi_i^2}
-4z_n\rho_0^2J\langle\cos{\phi}\rangle\cos{\phi_i}+2z_nJ\rho_0^2(1+\langle\cos{\phi}\rangle^2)\right)\psi_n(\phi_i)
= E_n\psi_n(\phi_i) . \label{eq:pH}
\end{equation}
This equation can be reduced to the standard Mathieu equation \cite{Mathieu}, using $v=\phi/2$, $y(v)=\psi_n(\phi_i/2)$, 
\begin{equation}
\frac{d^2y_n(v)}{dv^2}+(a_n-2q\cos{2v})y_n(v)=0 , \label{eq:mathieu_one}
\end{equation}
where the characteristic value of the Mathieu equation and $q$ are written as
\begin{eqnarray*}
a_n&=&4(E_n-2z_nJ\rho_0^2(1+\langle\cos{\phi}\rangle^2))\frac{2m\rho_0^2}{\hbar^2}=\frac{E_n-B(1+\langle\cos{\phi}\rangle^2)}{A} ,\\
q&=&-8z_n\rho_0^2J\langle\cos{\phi}\rangle\frac{2m\rho_0^2}{\hbar^2}=-\frac{B}{A}\langle\cos{\phi}\rangle ,
\end{eqnarray*}
where we define $A=\frac{\hbar^2}{8m\rho_0^2}$ and $B=2z_nJ\rho_0^2$. The eigenvalues are explicitly written as
\begin{equation}
E_n=Aa_n+B(1+\langle\cos{\phi}\rangle^2) .
\end{equation}
The allowed eigenfunctions are determined by the condition that the wave functions be single-valued, i.e., that $\psi_n(\phi+2\pi)=\psi_n(\phi)$, or equivalently, that $y_n(v+\pi)=y_n(v)$. The allowed three lowest solutions, up to the order of $q^2$, are \cite{Mathieu}
\begin{eqnarray*}
y_0(v,q)&=&\frac{1}{\sqrt{\pi}}\left[1-\frac{q}{2}\cos{2v}+q^2\left(\frac{\cos{4v}}{32}-\frac{1}{16}\right)\right], \quad a_0=-\frac{q^2}{2}, \\
y_2(v,q)&=&\frac{2}{\sqrt{\pi}}\left[\cos{2v}\!-\!q\left(\frac{\cos{4v}}{12}-\!\frac{1}{4}\right)\!+\!q^2\left(\frac{\cos{6v}}{384}-\frac{19\cos{2v}}{288}\right)\right],\quad\!\! a_2\!=\!4\!+\!\frac{5q^2}{12}, \\
y_{-2}(v,q)&=&\frac{2}{\sqrt{\pi}}\left[\sin{2v}-q\frac{\sin{4v}}{12}+q^2\left(\frac{\sin{6v}}{384}-\frac{\sin{2v}}{288}\right)\right], \quad a_{-2}=4-\frac{q^2}{12} ,
\end{eqnarray*}
where these are normalized like $\int_0^{2\pi}\psi_n(\phi)d\phi=1$. Thus, the matrix elements for $\cos{\theta}$ on the corresponding bases, $n=0,2,$ and $-2$, are
\begin{equation}
\langle\cos{\phi}\rangle=
\left( \begin{array}{ccc}
-\frac{q}{2} & \frac{1}{\sqrt{2}} & 0 \\
\frac{1}{\sqrt{2}} & \frac{5q}{12} & 0 \\
0 & 0 & -\frac{q}{12}
\end{array} \right).
\end{equation}

\begin{figure}
\begin{center}
\includegraphics[height=8cm,width=9cm]{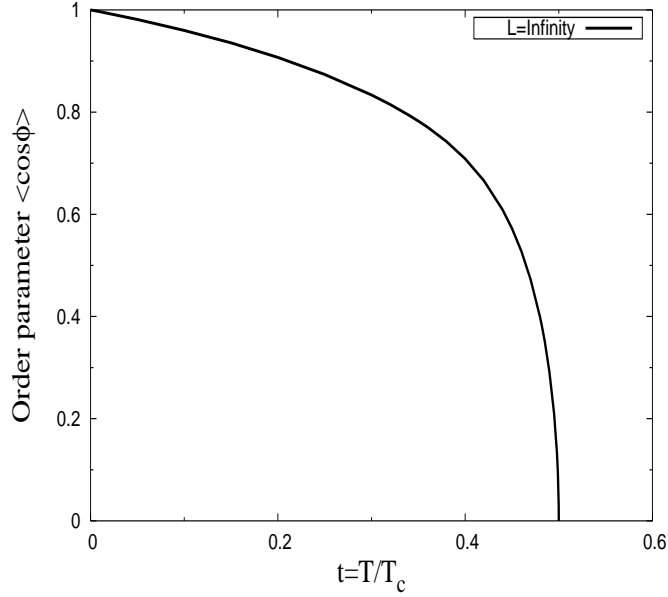}
\caption{\label{fig:order1} Temperature dependence of the phase-ordering for infinite long wires.
}
\end{center}
\end{figure}

From the mapping, we can get the self-consistent condition for the critical temperature of the phase ordering in terms of the parameters of the GL equation for sufficient or infinite long wires eq.\ (\ref{eq:gcondition}), which corresponds to the only consideration of the ground state $(n=0)$ in the quantum mechanics problem, and it takes the following form.
\begin{equation}
\langle\cos{\phi}\rangle=-\frac{A}{B}q = -\frac{q}{2}. \label{eq:conditionatt0}
\end{equation}
The temperature dependence of order parameter obtained by eq.\ (\ref{eq:conditionatt0}) for infinite long wires is shown in Fig.~\ref{fig:order1}.
This figure clearly shows that there is a second order phase transition at $t=0.5$ because the order parameter continuously becomes zero at the critical point. As expected, the critical temperature of the whole wires is lower than the critical temperature of a single wire.
The transition temperature of phase ordering can be calculated by finding the temperature where $\langle\cos{\phi}\rangle$ becomes zero. Thus, because with $ A\rightarrow\frac{m^*\xi_0^4}{8\hbar^2\beta^2\tilde{\Delta}^2(t)}$ and $B\rightarrow\frac{2z_nK\tilde{\Delta}^2(t)}{\xi_0^2}$,
\begin{equation}
\frac{B}{A}\rightarrow\frac{16z_n\hbar^2\beta^2K\tilde{\Delta}^4(t)}{m^*\xi_0^6}=2\alpha\frac{(1-t)^2}{t^2},
\end{equation}
where $\alpha=\frac{8z_n\hbar^2\tilde{\Delta}^4(0)K}{m^*\xi_0^6(k_BT_c^0)^2}$, the condition becomes
\begin{equation}
\langle\cos{\phi}\rangle = 0 \rightarrow t_c=\sqrt{\alpha}g^2(t_c) .
\end{equation}
Therefore, using $g(t)=\sqrt{1-t}$, this critical temperature becomes
\begin{equation}
t_c=\frac{\sqrt{\alpha}}{1+\sqrt{\alpha}} .
\end{equation}

\begin{figure}
\begin{center}
\includegraphics[height=8cm,width=9cm]{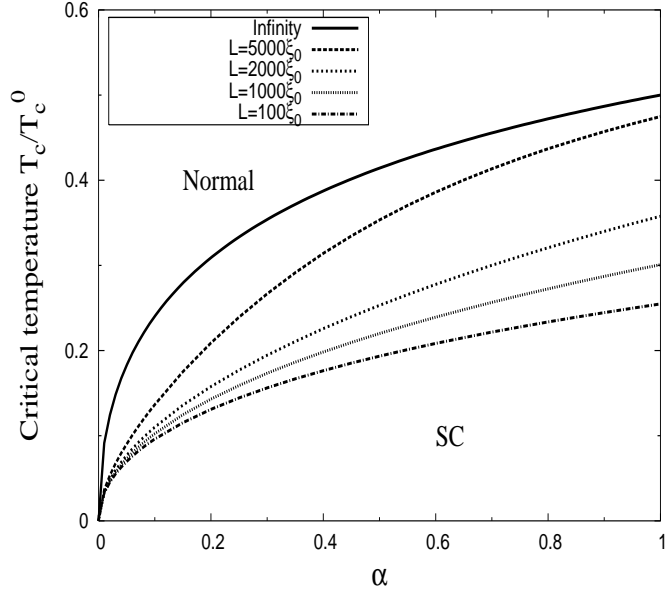}
\caption{\label{fig:phase1} Phase diagram of $t_c=T_c/T_c^0$ as a function of $\alpha$ for several values of length of the wires, $100\xi_0$,
$1000\xi_0$, $2000\xi_0$, $5000\xi_0$, and $\infty$ where $\xi_0=42$\AA .}
\end{center}
\end{figure}

On the other hand, for finite length wires, contributions from excited states in the quantum mechanics problem need to be considered because the effective temperature is not zero. Using up to the order $|n|\le 2$ for the solution of Mathieu's equation, using eq.\ (\ref{eq:condition}), the following self-consistent condition can be obtained,
\begin{equation}
-\frac{A}{B}q=\frac{-\frac{q}{2}e^{-\beta_{eff}E_0}+\frac{5q}{12}e^{-\beta_{eff}E_2}-\frac{q}{12}e^{-\beta_{eff}E_{-2}}}{e^{-\beta_{eff}E_0}+e^{-\beta_{eff}E_2}+e^{-\beta_{eff}E_{-2}}} ,
\end{equation}
where $\beta_{eff}E_n=\beta_{eff}(Aa_n+B(1+\langle\cos{\phi}\rangle))$, but the second term can be canceled. Therefore, with the mapping, we can get
\begin{equation}
\left(\frac{t}{1-t}\right)^2=\alpha \frac{1-\frac{2}{3}e^{-4x\frac{t}{1-t}}}{1+2e^{-4x\frac{t}{1-t}}},
\end{equation}
where we use the following mapping
\begin{equation}
\beta_{eff}A\rightarrow x\frac{t}{1-t}=x_0\frac{z_{max}}{\xi_0}\frac{t}{1-t}=\frac{m^*\xi_0^2k_BT_c^0}{8\hbar^2}\frac{\xi_0^2}{\tilde{\Delta}^2(0)}\frac{z_{max}}{\xi_0}\frac{t}{1-t} .
\end{equation}
Using the numerical values according to Tang {\it et al} \cite{Tang2001}, $x_0\approx 1.4\times 10^{-4}$. A plot of $T_c$ versus $\alpha$ for several lengths ($100\xi_0$,
$1000\xi_0$, $2000\xi_0$, and $5000\xi_0$) and infinite length are given in Fig.~\ref{fig:phase1}.
This figure shows that as the length of the wires has increased, the phase critical temperature has increased.

\subsection{Perpendicular magnetic field}
The critical temperature for the presence of a magnetic field on the wires can be obtained by solving the non-Hermitian eq.\ (\ref{eq:phaseonlyh}).
\begin{equation}
\left\{\!\!\frac{-\hbar^2}{2m\rho_0^2}\frac{\partial^2}{\partial\phi_i^2}
\!\!-\!\frac{e^*B_{eff}}{2mc}\frac{\hbar}{i}\frac{\partial}{\partial\phi_i}
\!\!-\!4z_n\rho_0^2J\langle\cos{\phi}\rangle\cos{\phi_i}\!\!+\!2z_nJ\rho_0^2(1\!\!+\!\langle\cos{\phi}\rangle^2)\!\right\}
\!\psi_n(\phi_i)
\!\!=\!E_n\psi_n(\phi_i). \label{eq:pAH}
\end{equation}
Using $\psi_n(\phi)=e^{pv}F(v)$ with and $v=\phi/2$ and
\begin{equation}
p=i\frac{2e^*\rho_0^2B_{eff}}{c\hbar},
\end{equation}
again this equation reduces to the standard Mathieu equation:
\begin{equation}
\frac{d^2F(v)}{dv^2}-(2q\cos{2v})F(v)=-a_{\nu}F(v),
\end{equation}
where
\begin{eqnarray*}
a_{\nu}-p^2&=&4(E_n-2z_nJ\rho_0^2(1+\langle\cos{\phi}\rangle^2))\frac{2m\rho_0^2}{\hbar^2}=\frac{E_n-B(1+\langle\cos{\phi}\rangle^2)}{A}, \\
q&=&-8z_n\rho_0^2J\langle\cos{\phi}\rangle\frac{2m\rho_0^2}{\hbar^2}=-\frac{B}{A}\langle\cos{\phi}\rangle .
\end{eqnarray*}

The allowed eigenvalues are determined by the boundary condition that $\psi_n(\phi+2\pi)=\psi_n(\phi)$, or equivalently $F(v+\pi)=\exp(-p\pi)F(v)$. Thus we are interested only in the Floquet solutions of the Mathieu equation with Floquet exponent $\nu= 2n+ip$, where $n=0,\pm1,\pm2,....$. These solutions are explicitly written as \cite{Mathieu}
\begin{equation}
F_\nu(v)=c_0e^{i\nu v}\left[1-q\left(\frac{e^{2iv}}{4(\nu+1)}-\frac{e^{-2iv}}{4(\nu-1)}\right)\right],
\end{equation}
where $c_0$ is a normalization constant.
The eigenvalues are, using $q=-\frac{B}{A}\cos{\theta}$,
\begin{equation}
a_{\nu}=\nu^2+\frac{q^2}{2(\nu^2-1)}.
\end{equation}
The allowed three lowest solutions, up to the order of $q^2$, are \cite{Mathieu}
\begin{eqnarray*}
\psi_{ip}(v)&\!\!\!\!\!=\!\!\!\!\!& \sqrt{\frac{1}{\pi}}\left(1-q\frac{\cos{2v}+p\sin{2v}}{2(1+p^2)}\right), \quad\!\!\!
  a_{ip}\!=\!-\frac{q^2}{2(1+p^2)},  \\
\psi_{2+ip}(v)&\!\!\!\!\!=\!\!\!\!\!& \sqrt{\frac{1}{\pi}}\left(e^{2iv}-\frac{q}{4}\left(\frac{e^{4iv}}{3+ip}-\frac{1}{1+ip}\right)\right), \quad\!\!\! a_{2+ip}\!=\!4(1+ip)+\frac{q^2}{2(-p^2+4ip+3)}, \\
\psi_{-2+ip}(v)&\!\!\!\!\!=\!\!\!\!\!& \sqrt{\frac{1}{\pi}}\left(e^{-2iv}+\frac{q}{4}\left(\frac{e^{-4iv}}{ip-3}-\frac{1}{ip-1}\right)\right), \quad\!\!\! a_{-2+ip}\!=\!4(1-ip)+\frac{q^2}{2(3-p^2-4ip)}.
\end{eqnarray*}
Left wave functions can be obtained from right wave function with $\psi_n^L(v,p)=\psi_n^R(v,-p)^*$.

\begin{figure}
\begin{center}
\includegraphics[height=8cm,width=9cm]{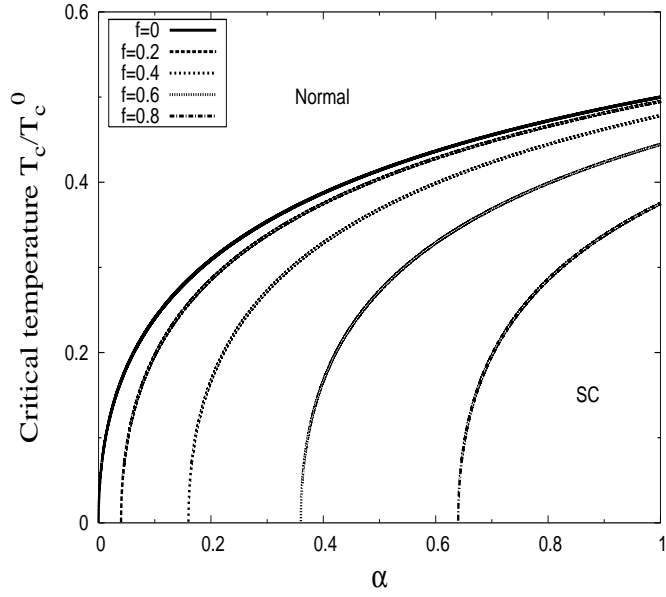}
\caption{\label{fig:phase2} Phase diagram of $t_c=T_c/T_c^0$ as a function of $\alpha$ for several values of magnetic field strength, $f=0.2$, $f=0.4$, $f=0.6$, and $f=0.8$ for infinity length wires}
\end{center}
\end{figure}

The self-consistent condition for long wires becomes the following form,
\begin{equation}
\langle \cos{\phi}\rangle=-\frac{A}{B}q = -\frac{q}{2(1+p^2)},
\end{equation}
because the matrix elements for $\cos{\theta}$ corresponding to $n=0,2,$ and $-2$ are, using $q=-\frac{B}{A}\langle\cos{\theta}\rangle$,
\begin{equation}
\langle\cos{\phi}\rangle=\frac{1}{2}
\left( \begin{array}{ccc}
-\frac{q}{1+p^2} & 1 & 1 \\
1 & \frac{q}{3+4ip-p^2} & \frac{1}{2(1+p^2)} \\
1 & \frac{1}{2(1+p^2)} & \frac{q}{3-4ip-p^2}
\end{array} \right) .
\end{equation} 
Again, we can determine the transition temperature of the phase ordering, where $\langle\cos{\phi}\rangle$ becomes zero.
\begin{equation}
\langle\cos{\phi}\rangle = 0 \rightarrow t_c=\sqrt{\frac{\alpha}{1+p^2(t_c)}}g^2(t_c).
\end{equation}
The approximation $g(t)=\sqrt{1-t}$ is again used for this case and then we can get
\begin{equation}
t_c=\frac{\sqrt{\alpha-f^2}}{1+\sqrt{\alpha-f^2}},
\end{equation}
where we define $f$ as
\begin{equation}
p\rightarrow\frac{fg^2(t)}{t}=f_0\frac{A_z\xi_0}{\Phi_0}\frac{g^2(t)}{t}=\frac{8\pi\hbar^2 }{k_BT_c^0m^*\xi_0^2}\frac{
\tilde{\Delta}^2(0)}{\xi_0^2}\frac{A_z\xi_0}{\Phi_0}\frac{1-t}{t},
\end{equation}
where $\Phi_0 = hc/e^*$. When $f=0$, this solution corresponds to the previous case. A plot
of $T_c$ versus $\alpha$ for $f=0, 0.2, 0.4, 0.6$, and $0.8$ is given in Fig.~\ref{fig:phase2}.
This figure shows that the critical temperatures have the minimum values for the interaction between the wires. These values can be calculated by
\begin{equation}
\alpha\geq f^2 \rightarrow z_nK \geq
\frac{8\pi^2\hbar^2}{m^*\xi_0^2}\left(\frac{A_z\xi_0}{\Phi_0}\right)^2.
\end{equation}

\begin{figure}
\begin{center}
\includegraphics[height=8cm,width=9cm]{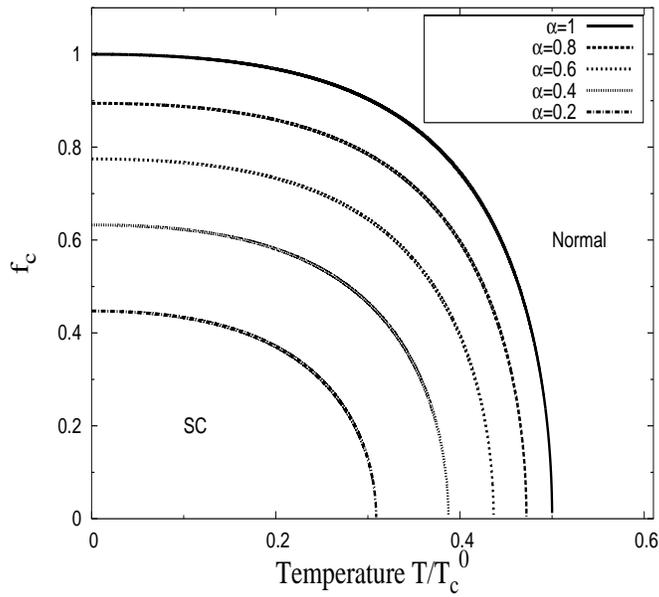}
\caption{\label{fig:phase3} Temperature dependence of the critical field strength, $f_c$ for $\alpha=0.2, 0.4, 0.6, 0.8$, and $1$ for infinity long wires.}
\end{center}
\end{figure}

The critical $f_c$, which is related to the maximum flux in the wires can be obtained,
\begin{equation}
f_c =\frac{\sqrt{\alpha (1-t)^2-t^2}}{1-t}.
\end{equation}
Near the critical temperature of phase ordering, using $t=\frac{\sqrt{\alpha}}{1+\sqrt{\alpha}}-\delta t$, this can be written
\begin{equation}
f_c\approx\sqrt{2}(1+\sqrt{\alpha})\alpha^{1/4}\sqrt{\delta t}.
\end{equation}
Fig.~\ref{fig:phase3} shows that this critical magnetic field $f_c$ for $\alpha=0.2, 0.4, 0.6, 0.8$, and $1$ as a function of a temperature.


\section{Summary}
We have presented a mapping between a one-dimensional GL problem in the presence of a vector potential along wires and a two-dimensional quantum mechanics problem with a perpendicular magnetic field. Moreover, in the case of weak links between wires, we have obtained, using the mean-field approximation, the phase diagrams for the presence of a magnetic field and absence of it.

Next, we discuss the parameters used in this paper. Using the numerical values of the various parameters appropriate to those of a single-walled carbon nanotube, which according to Tang {\it et al} \cite{Tang2001}, where superconducting with a relatively high transition temperature $T^0_c = 15 \mathrm{K}$, $k_BT^0_c = 1.3 \mathrm{meV}$, $\alpha_0T^0_c = 6 \mathrm{meV}$, $\gamma = 1.3 \mathrm{meV}$\AA,
$m^* = 0.36 m_{\mathrm{e}}$, and $\xi_0 = \frac{\hbar}{\sqrt{2 m^* \alpha_0 T^0_c}}= 42$\AA, we can obtain the following values for $\alpha$ and $f$, $\alpha=\frac{z_nK}{8.6\times 10^{-6}\mathrm{meV}}$ and $f=1.7\times 10^4\frac{A_z\xi_0}{\Phi_0}$. The Josephson coupling energy $K$ is approximated by $\frac{\hbar^2}{2m_cs^2}$ where $s$ is the distance between nearest wires. If we use $m_c=m^*$, $K$ can be written as $\frac{\xi_0^2\alpha_0 T^0_c}{s^2}$. Thus, supposing $s\approx 5$\AA, $K$ is order of $100\sim 1000 [\mathrm{meV}]$. Therefore, our values used in the figures are well suited for describing real systems.

We discuss about the use of the GL free energy functional. In principle, this free energy functional is applicable only near the critical temperature, $T-T_c^0\ll T_c^0$. Besides near the critical temperature $T_c^0$, the qualitative description of this functional may not be reasonable, although we can employ higher order expansions of the order parameter in the G.L equation.

We want to comment the effect on the interaction term by a magnetic field. When there is a magnetic field, the phase difference needs to be replaced by
$\phi_i-\phi_{i+1}-\frac{2\pi}{\Phi_0}\int\vec{A}\cdot d\vec{l}$ where the integration is between different wires. However, because the direction of vector potential is taken in the direction of the wires, $z$, there is no contribution from the integral on the phase difference.

In this paper, we only consider the periodic boundary condition for simplification. When wires are sufficient long, the effect of the boundary conditions may not change the physical properties of the system. However, these boundary conditions may affect the properties of the system because of finite length of wires. Moreover, our theory neglects the effects of disorder, which plays an important role on balk superconductors. With these degrees of freedom, the properties of the system may be changed. Thus, it might be an interest to consider these cases for our future research.

\end{document}